\documentclass[12pt]{article}

\usepackage{amsmath}
\usepackage{amsfonts}
\usepackage{amssymb}
\usepackage{amsthm}
\usepackage{setspace}
\usepackage{color}
\usepackage{subcaption}
\usepackage{caption}
\usepackage{enumerate}
\usepackage[hidelinks]{hyperref}
\usepackage{graphicx}
\usepackage[hidelinks]{hyperref} 
\usepackage[nottoc]{tocbibind}
\usepackage[utf8]{inputenc}

\usepackage{titlesec}

\titleformat*{\section}{\large\bfseries}

\newcommand\ee{\end{equation}}
\newcommand\be{\begin{equation}}
\newcommand\eea{\end{eqnarray}}
\newcommand\bea{\begin{eqnarray}}

\newcommand\mpl{M_{\rm pl}}
\newcommand\comment[1]{}

\comment{
\hypersetup{
colorlinks=true,
citecolor=DarkBlue,
linkcolor=DarkBlue,
urlcolor=DarkBlue,
}}

\def\O{\mathcal{O}}

\def\d{\partial}

\def\vphi{\varphi}

\def\rad{{\vphi}}
\def\pb{{\Delta S_{\rm p.b.}}}

\begin{document}

\begin{center}

  {\Large\bf An infrared bound on the ultraviolet bounce}

\vskip 1 cm
{Mehrdad Mirbabayi$^a$ and Giovanni Villadoro $^{a,b}$}
\vskip 0.5 cm

{\em $^a$ Abdus Salam International Centre for Theoretical Physics, Strada Costiera 11, 34151, Trieste, Italy}

{\em $^b$ INFN, Sezione di Trieste, Via Valerio 2, I-34127 Trieste, Italy}

\vskip 1cm

\end{center}
\noindent {\bf Abstract:} {\small Sometimes a local minimum is known to be a metastable vacuum inside the low-energy EFT, but the true vacuum lies outside, and the bounce solution mediating the decay cannot be found. For single-field decay, Espinosa has proposed a family of configurations called ``pseudo-bounces'' as a way to constrain the decay rate. They are parametrized by the central field value $\Phi_0$ and constructed without the knowledge of the true vacuum. We prove that the pseudo-bounce family has a monotonically decreasing decay exponent whose end point and minimum is the bounce action, and this continues to hold when gravitational effects are non-negligible. We then use this to estimate the decay rate (in the radion direction) of the promised AdS$_3$ vacuum of the Standard Model. }
\vskip 1 cm

\section{Introduction}

Finding the $O(D)$-symmetric bounce solution, $D$ counting the number of spacetime dimensions, is the standard method for calculating the decay rate of a metastable vacuum \cite{Coleman}. However, if all we know is the effective theory built around the metastable minimum, the decay might be UV dominated and uncalculable. As a rule of thumb, this is expected if there is a field direction along which the potential falls faster than $-\Phi^{2D/(D-2)}$, to be turned around at some unknown $\Phi_{\rm UV}$ due to some unknown UV physics. The Higgs potential might be a marginal example of this situation \cite{SM}. The Casimir-energy supported AdS$_3$ vacuum of the Standard Model \cite{Arkani} is another example \cite{Shiu}. 

In such cases, it would be useful to find an estimate of the decay rate -- a {\em lower bound} as we expect a faster rate the higher the UV scale. Such a lower bound can indeed be obtained using configurations called ``pseudo-bounces'', assuming the tunneling is captured by a single scalar field with the Euclidean action
\be\label{S}
S = \int d^D x \left[\frac{1}{2} (\nabla\Phi)^2 + U(\Phi)\right].
\ee
Pseudo-bounces have been introduced in \cite{Espinosa} and generalized to the gravitational case in \cite{Espinosa_gr}. A convenient way to describe them is as a generalization of the $O(D)$-symmetric bounce solution: $O(D)$ symmetry simplifies the field equation to
\be\label{ON}
\Phi'' + \frac{D-1}{\xi} \Phi' = \frac{dU}{d\Phi},
\ee
where $\xi = |x^i|$ and prime denotes derivative with respect to it. The above equation can be thought of as describing the motion of a particle in the potential $-U$, and with a ``time''-dependent friction coefficient $(D-1)/\xi$. The bounce solution is a solution that at $\xi=0$ is at rest, with positive ``energy'', $-U(\Phi(0))>0$, and asymptotically arrives at the false vacuum $\Phi(\xi\to \infty) = 0$, where $U(0) =0$. Coleman proves the existence of such a solution by an undershoot-overshoot argument. All regular solutions of \eqref{ON} have $\d_\xi\Phi(0) = 0$. Those with $\Phi(0)$ too close to the false vacuum, won't reach it because they do not have enough energy. The ones that start too close to the true vacuum [where $-U(\Phi_{\rm TV}) >0, U'(\Phi_{\rm TV}) = 0$] spend too long near it, will experience insufficient friction and overshoot. By continuity, a bounce solution must exist at the boundary of the two regimes.

Certainly, trajectories that remain close to the false minimum are realizable in an EFT that is built around that minimum. The absence of a bounce in this EFT is because all such trajectories that are {\em regular solutions} undershoot. They have infinite action and therefore do not give a useful bound.

{\em Definition.} The ``pseudo-bounces'' are a family of $C^1$ functions parametrized by a single number $\Phi_0$ and defined by $\Phi(\xi\leq \xi_0)=\Phi_0$, $\Phi(\xi\to \infty) = 0$, and a solution of \eqref{ON} for $\xi_0<\xi<\infty$. These conditions implicitly fix $\xi_0$ in terms of $\Phi_0$, and the range of $\Phi_0$ by the requirement that $\xi_0\geq 0$. If $-U(\Phi_0)>0$ and the regular solution of \eqref{ON} with $\Phi(0) = \Phi_0$ undershoots, there must exist a pseudo-bounce with finite $\xi_0>0$. Long enough suspense leads to overshooting. The bounce solution lives at the (possibly infinitely far) boundary of the pseudo-bounce family, where $\xi_0(\Phi_0) =0$.

Heuristically, each pseudo-bounce describes tunneling to a configuration with a bubble of size $\xi_0$ and fixed $\Phi= \Phi_0$ in the interior. It is the bounce solution of a modified potential that coincides with $U$ before reaching $\Phi_0$, then sharply, but continuously, turns around. 

\section{Monotonicity of the pseudo-bounce action without gravity}
Every pseudo-bounce gives an upper bound on the bounce action. One proof of this (due to Espinosa \cite{Espinosa}) relies on the fact that by formulating the QFT problem as a multi-variable quantum mechanics problem, pseudo-bounce is a path from the false vacuum to a zero-energy, static configuration (which is the symmetric slice of the pseudo-bounce). Among such paths bounce has the minimum action \cite{Coleman_negative}. Another proof can be given via a simple generalization of Coleman-Glaser-Martin theorem \cite{CGM}. Neither of these arguments apply to the gravitational case.

The pseudo-bounce action has also been observed to have a monotonic behavior, decreasing toward the bounce endpoint \cite{Espinosa_rev}. We will give a proof of this stronger result, which can be generalized to the gravitational case. Pseudo-bounce is composed of two pieces. Hence, its action can be decomposed as $S_{\rm p.b.}= S_1 +S_2$, with
\be\begin{split}
S_1  &=\frac{\Omega_{D-1}}{D} \xi_0^D U(\Phi_0),\\[10pt]
S_2 & = \Omega_{D-1} \int_{\xi_0}^\infty d\xi \xi^{D-1} \left[ \frac{1}{2} {\Phi'}^2 + U(\Phi)\right],\end{split}\ee
where $\Omega_{D-1}$ is the volume of $D-1$ unit sphere, and $\xi_0$ is implicitly fixed as a function of $\Phi_0$. Taking the derivative with respect to $\Phi_0$, we get
\be\label{dS1}
\frac{d S_1(\xi_0(\Phi_0),\Phi_0)}{d\Phi_0} = \Omega_{D-1}\frac{\xi_0^D}{D}\frac{dU(\Phi_0)}{d\Phi_0}
+ \Omega_{D-1} \frac{d\xi_0}{d\Phi_0} \xi_0^{D-1} U(\Phi_0).
\ee
For $S_2$, we use the fact that for $O(D)$ symmetric configurations our system has reduced to a mechanics problem with one degree of freedom $\Phi$ and ``time'' $\xi$. $S_2$ is the on-shell action for motion from $\Phi_0$ at $\xi_0$ to $\Phi=0$ at $\xi = \infty$. Partial derivative of $S_2$ with respect to $\Phi_0$ and $\xi_0$ are related to the conjugate momentum and Hamiltonian of this mechanical system
\be\begin{split}
\d_{\Phi_0} S_2 &= -\Omega_{D-1} \xi^{D-1} \Phi'_0,\\
\d_{\xi_0} S_2 & =  \Omega_{D-1} \xi^{D-1} \left[ \frac{1}{2} {\Phi_0'}^2 - U(\Phi_0)\right].\end{split}
\ee
On the pseudo-bounce $\Phi'_0 \equiv \Phi'(\xi_0) =0$ and hence
\be\label{dS2}
\frac{d S_2(\xi_0(\Phi_0),\Phi_0)}{d\Phi_0} = - \Omega_{D-1} \frac{d\xi_0}{d\Phi_0} \xi_0^{D-1} U(\Phi_0).
\ee
In total, we find
\be\label{mono}
\frac{dS_{\rm p.b.}(\Phi_0)}{d\Phi_0} =\Omega_{D-1} \frac{\xi_0^D}{D}\frac{dU(\Phi_0)}{d\Phi_0}.
\ee
$S_{\rm p.b.}$ decreases as $U(\Phi_0)$ decreases, becoming stationary at the bounce, where $\xi_0 =0$.

\section{Monotonicity of the pseudo-bounce action with gravity}
In the gravitational setup, the absolute value of the vacuum energy affects the decay rate \cite{CDL}. We first restrict to the case where false vacuum energy is non-positive, i.e. $U(0)\leq 0$. A slight modification of the argument leads to the same conclusion for dS false vacua. We continue restricting to $O(D)$ symmetric configurations. Now in addition to the scalar field there is one metric degree of freedom $\rho$:
\be
ds^2 = d\xi^2 + \rho(\xi)^2 d\Omega^2,
\ee
where $d\Omega^2$ is the line element of a unit $S^{D-1}$. Up to boundary terms that either vanish, or cancel in the differences, the Einstein-Hilbert plus scalar field action reads
\be
S = \Omega_{D-1} \int d\xi \left[\rho^{D-1}\left(\frac{1}{2} {\Phi'}^2+U\right)
  -\frac{1}{\tilde\kappa} \rho^{D-3} (1+{\rho'}^2)\right],
\ee
where in terms of the conventional gravitational coupling $\tilde \kappa\equiv \frac{2\kappa}{(D-1)(D-2)}$. Having set the lapse function to $1$, we have to impose the Hamiltonian constraint by hand
\be\label{Fr}
-{\rho'}^2+ 1 + \tilde\kappa \rho^2 \left(\frac{1}{2} {\Phi'}^2-U\right) =0.
\ee
The scalar field equation is modified to
\be\label{phi}
\Phi''+(D-1) \frac{\rho'}{\rho} \Phi' = \frac{dU}{d\Phi}.
\ee
Pseudo-bounces are defined as before, except during the period of suspense $0\leq \xi\leq \xi_0$, we solve \eqref{Fr} for $\rho$. By a change of integration variable and using \eqref{Fr}, the action in the suspense period becomes
\be
S_1 = -\frac{2\Omega_{D-1}}{\tilde \kappa}\int_0^{\rho_0} d\rho \rho^{D-3} \sqrt{1-\tilde \kappa\rho^2 U(\Phi_0)}.
\ee
It follows that
\be\label{dS1_gr}\begin{split}
\d_{\rho_0} S_1 &= - \frac{2\Omega_{D-1}}{\tilde\kappa}\rho_0^{D-3}\sqrt{1-\tilde\kappa\rho_0^2 U(\Phi_0)},\\[10 pt]
\d_{\Phi_0} S_1 &= \frac{d U(\Phi_0)}{d\Phi_0} \Omega_{D-1}  \int_0^{\rho_0} \frac{d\rho \rho^{D-1}}{\sqrt{1-\tilde\kappa\rho^2 U(\Phi_0)}}.\end{split}
\ee
Similarly, the action of the trivial saddle is
\be
S_{\rm FV}= -\frac{2\Omega_{D-1}}{\tilde\kappa}\int_0^{\infty} d\rho \rho^{D-3} \sqrt{1-\tilde\kappa\rho^2 U(0)}.
\ee
Both this and the rolling part of the pseudo-bounce action ($S_2$) diverge. We are interested in the difference $\Delta S_{\rm p.b.}=S_1+S_2 -S_{\rm FV}$. The rolling part is a solution to the equations of motion with initial conditions $\rho_0,\Phi_0$ at ``time'' $\xi_0$ and final conditions $\rho=\infty,\Phi=0$ at ``time'' $\xi= \infty$. Derivatives of the on-shell action with respect to the initial conditions at fixed final conditions are related to the conjugate momenta and Hamiltonian via
\be\begin{split}
\d_{\Phi_0}S_2 &= -\Omega_{D-1}\rho_0^{D-1} \Phi'_0,\\[10pt]
\d_{\rho_0}S_2 &= \frac{2\Omega_{D-1}}{\tilde\kappa}\rho^{D-3}_0\rho'_0,\\[10pt]
\d_{\xi_0}S_2 &= \frac{\Omega_{D-1}}{\tilde\kappa} \rho_0^{D-3} \times (\text{LHS of \eqref{Fr}}).\end{split}
\ee
On the pseudo-bounce $\xi_0$ and $\rho_0$ are determined in terms of $\Phi_0$. Since $\Phi'(\xi_0)=0$ and \eqref{Fr} is always satisfied, only $\d_{\rho_0}S_2\neq 0$ among the above three. It cancels with the first line of \eqref{dS1_gr} when computing $d\Delta S_{\rm p.b.}/d\Phi_0$, and yields
\be\label{dpb/df0}
\frac{d \Delta S_{\rm p.b.}}{d\Phi_0} =\frac{d U(\Phi_0)}{d\Phi_0} \Omega_{D-1}\int_0^{\rho_0} \frac{d\rho \rho^{D-1}}{\sqrt{1-\tilde\kappa\rho^2 U(\Phi_0)}},
\ee
which implies that $\Delta S_{\rm p.b.}$ decreases as $U(\Phi_0)$ decreases.

Unlike the non-gravitational case the existence of pseudo-bounces with finite $\Delta S_{\rm p.b.}$ are not guaranteed. If a bounce exists, then it will be the end point of an interval of pseudo-bounces since it corresponds to a pseudo-bounce with zero suspense. But as $\Phi_0$ climbs toward the false vacuum, this interval ends at a threshold $\Phi_{\rm th}$ with $U(\Phi_{\rm th})<0$ at which $\Delta S_{\rm p.b.}(\Phi_{\rm th}) = \infty$.

If the false vacuum has positive vacuum energy, there is always the possibility of a thermal (Hawking-Moss \cite{HM}) transition, whose decay exponent can be bound by the de Sitter entropy. Whether or not a distinct Coleman-De Luccia bounce \cite{CDL} exists or if it dominates depends on the potential. The bounce and pseudo-bounces must now satisfy a different final condition since $\rho$ reaches a maximum and then contracts to zero at some finite $\xi_f$. For regularity, one demands $\Phi'_f = 0$ but the final value $\Phi_f$ is not fixed. It depends on $\Phi_0$. In the derivation of Eq. \eqref{dpb/df0}, this adds an extra piece $\frac{d\Phi_f}{d\Phi_0} \d_{\Phi_f}S_2$, which however vanishes because $\d_{\Phi_f}S_2 \propto \Phi'_f = 0$.

\section{Application: AdS$_3$ vacuum of Standard Model}
For some experimentally allowed choices of neutrino masses, the Standard Model has an AdS$_3$ vacuum when compactified on a circle \cite{Arkani}. The size of the circle (radion) is stabilized by a balance between the negative Casimir energy of photons and gravitons and the positive Casimir energy of neutrinos. This is a metastable vacuum because when the size of the compact dimension becomes comparable to the mass of charged particles, there will be choices of the gauge-field Wilson lines for which the net Casimir contribution becomes negative and radion potential decreases exponentially within the Standard Model \cite{Shiu}. This exponential fall off will presumably stop, if not earlier, when the compact dimension becomes of Planckian size.

We can use pseudo-bounces to put a bound on the decay rate of the AdS$_3$ vacuum, at least to make sure that within the Standard Model it has longer lifetime than its curvature length. This is an effectively 1-dimensional problem, with kinetic and potential terms
\be\label{kin}
T = \frac{1}{\kappa_3} (\d\rad)^2, \qquad U = \frac{f(\rad)}{ r_\nu^3}e^{6\rad},
\ee
where $2\pi r_\nu\sim \frac{1}{m_\nu}$ is the size of the compact dimension at the metastable vacuum, $\rad$ is the dimensionless radion field $\rad =\log\frac{r_\nu}{r}$, and the $3d$ gravitational coupling is related to $4d$ parameters via $\kappa_3 = \frac{4 G_N}{r_\nu}=\frac{1}{2\pi r_\nu \mpl^2}$. The function $f(\rad)$ is almost constant except for rapid jumps at mass-thresholds. For instance, for Majorana neutrinos $f(\rad) \approx 9 \times 10^{-4}$ in the range $m_\nu \ll \frac{1}{2\pi r}\ll m_e$ and $f(\rad)\approx - 0.03$ for $m_{\rm top}\ll \frac{1}{2\pi r}$ up to the next threshold beyond the Standard Model. 

Given the monotonicity of the pseudo-bounce action as a function of the central value $\rad_0$, the strongest bound comes from the highest known energy scale, namely $\rad_0 >\log\frac{m_{\rm top}}{m_\nu} \sim 30$. We will find the leading large $\rad_0$ behavior of $\pb$. 

For an $O(3)$ symmetric configuration the equations of motion read
\be\begin{split}
{\rho'}^2 &= 1+ \rho^2({\rad'}^2-\kappa_3 U),\\[10pt]
\rad'' & = -\frac{2\rho'}{\rho}\rad' + \kappa_3 \frac{d U}{d\rad}.
\end{split}\ee
Gravitational effects are inevitably important for a decay in radion direction since its interactions are of gravitational strength as follows from \eqref{kin}. In particular, during the suspense period where $\rad' =0$ and only the energy constraint is satisfied the $\kappa_3 U(\rad_0)$ term cannot be neglected. Indeed, as we will shortly verify, by the end of the suspense period the potential energy dominates the curvature term and leads to $\rho' \gg 1$. Soon after the suspense period though the potential becomes irrelevant because of its sharp $\rad$ dependence and the motion becomes kinetic energy dominated. Integrating once gives
\be
\rad' \approx -\eta\sqrt{-\kappa_3 U(\rad_0)} \frac{\rho_0^2}{\rho^2},\qquad \rho' \approx - \rho \rad',
\ee
where the integration constant $\eta$ would be $1$ if at $\rho_0$ the potential energy were instantaneously converted to kinetic energy. In practice, this takes of order of a Hubble time and $\eta \approx 3$. Integrating once more, we find
\be
\rad-\rad_0 \approx \log\frac{\rho_0}{\rho}.
\ee
As a consistency check, this implies $U(\rad)\approx U(\rad_0)\frac{f(\rad)}{f(\rad_0)}\frac{\rho_0^6}{\rho^6}$, which for $\rho\gg \rho_0$ remains much smaller than the kinetic energy in magnitude.

As in FLRW cosmology, kinetic energy dilutes faster than spatial curvature and its domination ends at $\rho_1$ such that
\be\label{rho1}
\rho_1^2 {\rad'}^2 =1 \Rightarrow \rho_1 \approx \eta \sqrt{-\kappa_3 U(\rad_0)} \rho_0^2.
\ee
Afterward, $\rho' \approx 1$ and $\rad$ will roll for an $\O(1)$ amount. Hence a successful pseudo-bounce, namely that $\rad$ arrives to $0$ as $\rho\to \infty$, implies $\rad_0 \approx \log\frac{\rho_1}{\rho_0}$ and combined with \eqref{rho1}
\be
\rho_0 \approx \frac{1}{\eta \sqrt{-\kappa_3 U(\rad_0)}}e^{\rad_0} \propto e^{-2 \rad_0}.
\ee
Using this, we can estimate that $\pb \propto e^{-\rad_0}$ at large $\rad_0$. The proportionality coefficient can be determined using \eqref{dpb/df0}. Indeed, for $-\kappa_3 U(\rad_0)\rho_0^2\gg 1$
\be
\frac{d\pb}{d\vphi_0} \approx \frac{2\pi}{\sqrt{-\kappa_3 U(\rad_0)}}\rho_0^2 \frac{d U(\rad_0)}{d\rad_0}
=-\frac{12\pi}{\eta^2} \sqrt{\frac{r_\nu^3}{\kappa_3^3 f(\vphi_0)}} e^{-\rad_0}.
\ee
Integrating this and writing the result in terms of energy scales $\mpl, m_\nu$ and the UV cutoff $\Lambda \sim m_\nu e^{\rad_0}$, we find
\be\label{SLambda}
\pb(\Lambda) \sim \frac{\mpl^3}{m_\nu^2 \Lambda}.
\ee
We have argued that the true bounce is the end point of the family of pseudo-bounces. Interestingly, even if this end-point is at $\Lambda \sim \mpl$, the bounce action is $\sim \frac{\mpl^2}{m_\nu^2}\gg 1$. It is well-known, at least in the thin-wall regime \cite{CDL}, that gravitational effects can forbid the decay from a higher minimum with non-positive vacuum energy to a lower minimum. Here, we see an example where the decay exponent is still finite but parametricaly larger than what a non-gravitational analysis would suggest: Assuming $\rho = \xi$, the suspense period $\xi_0$ can be estimated by balancing the kinetic and the potential terms, leading to 
\be
\xi_0 \sim \frac{\mpl}{m_\nu^2}e^{-3\vphi_0},\qquad \text{(rigid $3d$ geometry)}
\ee
and a decay exponent $B \sim \frac{\mpl^3}{\Lambda^3}$. 

The above example is a counter-example to the proposed upper bound in \cite{Brown}, which predicts $B<\O(\mpl^3/\Lambda^3)$. Nevertheless, when gravitational effects are negligible, the proposal of \cite{Brown} is an efficient and reliable method for bounding the decay exponent, and it is applicable to pseudo-bounces via the generalization in their Eq. (32).

\section{Conclusion}

If a vacuum can be diagnosed as being metastable within a low-energy EFT, pseudo-bounces are useful for computing an upper bound on the decay exponent when the true vacuum and the bounce solution lie outside the EFT. Our argument holds both with and without gravity, however it assumes that a {\em single} scalar field interpolates between the two vacua, which is an assumption about the UV physics. In the non-gravitational case, other arguments guarantee that any pseudo-bounce gives a lower bound to the bounce action, but we are not aware of a generalization of those to the gravitational case. It would be interesting to explore the generalization of our result to multi-field scenarios. 

We used our single-field strategy to estimate the decay rate of the putative AdS$_3$ vacuum of Standard Model compactified on a circle. We found that because of the gravitational effects a hypothetical decay in the radion direction is exponentially slow even if the circle size at the true vacuum is Planckian. Of course, we cannot rule out other, faster decay channels beyond the Standard Model.

\vspace{0.3cm}
\noindent
\section*{Acknowledgments}

We thank Oliver Janssen for useful discussions.

\bibliography{bibdecay}

\providecommand{\href}[2]{#2}\begingroup\raggedright\begin{thebibliography}{10}

\bibitem{Coleman}
S.~R. Coleman, ``{The Fate of the False Vacuum. 1. Semiclassical Theory},''
  \href{http://dx.doi.org/10.1103/PhysRevD.16.1248}{{\em Phys. Rev. D}
  {\bfseries 15} (1977) 2929--2936}. [Erratum: Phys.Rev.D 16, 1248 (1977)].

\bibitem{SM}
D.~Buttazzo, G.~Degrassi, P.~P. Giardino, G.~F. Giudice, F.~Sala, A.~Salvio,
  and A.~Strumia, ``{Investigating the near-criticality of the Higgs boson},''
  \href{http://dx.doi.org/10.1007/JHEP12(2013)089}{{\em JHEP} {\bfseries 12}
  (2013) 089}, \href{http://arxiv.org/abs/1307.3536}{{\ttfamily arXiv:1307.3536
  [hep-ph]}}.

\bibitem{Arkani}
N.~Arkani-Hamed, S.~Dubovsky, A.~Nicolis, and G.~Villadoro, ``{Quantum Horizons
  of the Standard Model Landscape},''
  \href{http://dx.doi.org/10.1088/1126-6708/2007/06/078}{{\em JHEP} {\bfseries
  06} (2007) 078}, \href{http://arxiv.org/abs/hep-th/0703067}{{\ttfamily
  arXiv:hep-th/0703067}}.

\bibitem{Shiu}
Y.~Hamada and G.~Shiu, ``{Weak Gravity Conjecture, Multiple Point Principle and
  the Standard Model Landscape},''
  \href{http://dx.doi.org/10.1007/JHEP11(2017)043}{{\em JHEP} {\bfseries 11}
  (2017) 043}, \href{http://arxiv.org/abs/1707.06326}{{\ttfamily
  arXiv:1707.06326 [hep-th]}}.

\bibitem{Espinosa}
J.~R. Espinosa, ``{A Fresh Look at the Calculation of Tunneling Actions},''
  \href{http://dx.doi.org/10.1088/1475-7516/2018/07/036}{{\em JCAP} {\bfseries
  07} (2018) 036}, \href{http://arxiv.org/abs/1805.03680}{{\ttfamily
  arXiv:1805.03680 [hep-th]}}.

\bibitem{Espinosa_gr}
J.~R. Espinosa, ``{Fresh look at the calculation of tunneling actions including
  gravitational effects},''
  \href{http://dx.doi.org/10.1103/PhysRevD.100.104007}{{\em Phys. Rev. D}
  {\bfseries 100} no.~10, (2019) 104007},
  \href{http://arxiv.org/abs/1808.00420}{{\ttfamily arXiv:1808.00420
  [hep-th]}}.

\bibitem{Coleman_negative}
S.~R. Coleman, ``{Quantum Tunneling and Negative Eigenvalues},''
  \href{http://dx.doi.org/10.1016/0550-3213(88)90308-2}{{\em Nucl. Phys. B}
  {\bfseries 298} (1988) 178--186}.

\bibitem{CGM}
S.~R. Coleman, V.~Glaser, and A.~Martin, ``{Action Minima Among Solutions to a
  Class of Euclidean Scalar Field Equations},''
  \href{http://dx.doi.org/10.1007/BF01609421}{{\em Commun. Math. Phys.}
  {\bfseries 58} (1978) 211--221}.

\bibitem{Espinosa_rev}
J.~R. Espinosa, ``{A Unified View of Vacuum Decay Channels},'' in {\em {35th
  Rencontres de Blois}: {Particle Physics and Cosmology}}.
\newblock 1, 2025.
\newblock \href{http://arxiv.org/abs/2501.09714}{{\ttfamily arXiv:2501.09714
  [hep-th]}}.

\bibitem{CDL}
S.~R. Coleman and F.~De~Luccia, ``{Gravitational Effects on and of Vacuum
  Decay},'' \href{http://dx.doi.org/10.1103/PhysRevD.21.3305}{{\em Phys. Rev.
  D} {\bfseries 21} (1980) 3305}.

\bibitem{HM}
S.~W. Hawking and I.~G. Moss, ``{Supercooled Phase Transitions in the Very
  Early Universe},'' \href{http://dx.doi.org/10.1016/0370-2693(82)90946-7}{{\em
  Phys. Lett. B} {\bfseries 110} (1982) 35--38}.

\bibitem{Brown}
A.~R. Brown, ``{Thin-wall approximation in vacuum decay: A lemma},''
  \href{http://dx.doi.org/10.1103/PhysRevD.97.105002}{{\em Phys. Rev. D}
  {\bfseries 97} no.~10, (2018) 105002},
  \href{http://arxiv.org/abs/1711.07712}{{\ttfamily arXiv:1711.07712
  [hep-th]}}.

\end{thebibliography}\endgroup
\end{document}